\documentclass[a4paper]{article}

\usepackage[pages=none, color=white, position={current page.south}, placement=bottom, scale=1, opacity=1, vshift=5mm]{background}

\usepackage[margin=1in]{geometry} 

\usepackage{url}

\usepackage{amsmath}
\usepackage{amsthm}
\usepackage{amssymb}

\usepackage[utf8]{inputenc}
\usepackage{hyperref}
\hypersetup{
	unicode,
	pdfauthor={Author One, Author Two, Author Three},
	pdftitle={A simple article template},
	pdfsubject={A simple article template},
	pdfkeywords={article, template, simple},
	pdfproducer={LaTeX},
	pdfcreator={pdflatex}
}

\usepackage[numbers]{natbib}
\theoremstyle{plain}

\theoremstyle{definition}

\usepackage{graphicx, color}
\graphicspath{{fig/}}

\usepackage{algorithm, algpseudocode} 
\usepackage{mathrsfs} 

\usepackage{lipsum}

\title{Fragmented digital connectivity and security at sea}
\author{Rikke Bjerg Jensen\\
Information Security Group\\
Royal Holloway, University of London\\
rikke.jensen@rhul.ac.uk}
\date{
}

\begin{document}
	\maketitle
	
\begin{abstract}
\noindent This paper explores how uneven and often unreliable digital connections shape the patterns and routines of everyday life, work and rest for seafarers, during long periods at sea. Such fragmented connections, which surface when the ship moves in and out of connectivity or when onboard data allowances run out, create a series of uncertainties that might unsettle individual and collective notions of security. Ethnographic in nature, the study engaged 43 seafarers on board two container ships in European waters, during two two-week voyages between February and April 2018. This provided an empirically grounded exploration of how digitally facilitated connections, relations and networks, enabled through increasingly connected ships, shape and reshape seafarer lives. Findings from this study demonstrate the creative ways in which seafarers navigate and negotiate digitally facilitated connections to maintain relational ties with family and friends. The paper concludes by setting out future research directions and practical implications that speak to connectivity and security at sea.

\paragraph{Keywords:} digital technology; security; seafarers; ethnography; social relations 
\end{abstract}

 \section{Introduction}\label{sec:intro}
The maritime industry has been slow to adopt digitally enabled technology across the world's commercial fleet, limiting seafarers' ability to connect with their family and friends during periods at sea. Because of this, seafarers often live and work for months at a time in small international crews, while having little contact with life at home. This study engages with the growing reliance on digital technology within the seafaring community on the one hand and the fragmentation of this digitalisation on the other. In so doing, it aims to understand how fragmented connections shape social relations and interactions as well as routines and patterns of practice and, in turn, notions of both security and insecurity among seafarers. It brings to light the extent to which routines and practices are disrupted as a result of uneven and unreliable digital connectivity, and the implications this has for security.

The importance that seafarers attach to having access to digital technology has been reported in a series of industry reports and publications, e.g.~\cite{Futurenautics18,Nautilus17,SIRC:SEATT18}. It is also evident from existing work that seafarers often rely on digitally enabled access to remain connected with people at home during long periods at sea~\cite{MissiontoSeafarers19}. Moreover, the extent to which seafarers and their social networks use digital technology has been documented in a series of industry reports, often based on findings from survey data, e.g.~\cite{Futurenautics18,HSBA18,MissiontoSeafarers19,Nautilus17,SIRC:SEATT18}, as further detailed in Section~\ref{sec:related-work}. The present study is situated in the seafarer contexts that underpin such reports. While it focuses on digital connectivity onboard ships, it situates the technology within the relational and social assemblages that support and are supported through digital connections. In so doing, it ties into an emergent body of scholarship that sits at the intersections of Human-Computer Interaction (HCI), individual and collective security practices and ground-up research engagements, see e.g.~\cite{HAI:ColHan17,CHI:ColNew18,ColJen19,SP:SLIRK18,CHI:WenJenCol20}. 

The study is based on ethnographic fieldwork with 43 seafarers onboard two container ships during two two-week voyages in European waters in February and April 2018, grounding it in the daily practices and routines of seafarer lives. While several studies have used surveys to try to establish the rate of digital connectivity and its wide-ranging implications, e.g.~\cite{HSBA18,Futurenautics18,Nautilus17,MissiontoSeafarers19}, this study does not focus on the state of digital connectivity and what seafarers do with it. Instead, it attends to how and why seafarers navigate and negotiate a web of digital connections -- and the meanings they ascribe to their experiences of doing so. It foregrounds the mundane, social and lived experiences of seafarers' use of digitally enabled technology to build and maintain daily routines and established practices. By not separating technology from the social relations in which it is embedded, the study unpicks how uneven digital connections also disrupt such routines. More generally, the study unearths the effects of connectivity, no connectivity, poor connectivity, and their combining, on seafarers' own notions of security. This is directly linked to their sense of wellbeing, crew cohesion and social isolation and exclusion. Within this framing, the study presents the following questions for security research: \emph{how does fragmented digital connectivity onboard container ships impact on seafarers' relational ties and social interactions? And how does this influence seafarers' notions of security?} 
  
In responding to such questions, the study contributes to practice and policy discussions on the integration of digital technology onboard ships by providing an empirically driven and emerging research agenda. This is important since unreliable and uneven digital connectivity on the one hand and notions of security on the other are closely related for seafarers in their daily patterns of practice. These notions are often amplified through the confinements of the ship, while they are managed through both digitally enabled and offline routine activities. 

\section{Grounding Seafaring in Digital and Security Scholarship}\label{sec:related-work}
This section focuses on three broader areas of scholarship. First, Section~\ref{sec:communicating-at-sea} discusses existing scholarship and industry reports pertaining to digital connectivity (or lack thereof) onboard ships. Second, Section~\ref{sec:digital-exclusion} engages with HCI research on notions of digital exclusion, particularly within marginalised groups where an understanding of security as grounded in the social underpinnings of technology has been employed. Finally, Section~\ref{sec:notions-security} brings to light an interpretation of security that takes into account the lived experiences of security. This section grounds research findings in these distinct yet overlapping bodies of scholarship. In so doing, it highlights how the limitations of existing work on digital connectivity at sea can be both expanded and deepened by engaging with situated HCI research and notions of security that engage with their social contexts. 

This is relevant for the present study as it employs an understanding of security decoupled from a specific technological focus by exploring the social interactions and relations facilitated through digital technology -- and the security that such engagements enable or, indeed, hinder. To this end, it bridges specific notions of security with HCI scholarship in order to establish a backdrop against which the findings of this study can be positioned and as a lens through with onboard digital practices can be explored. While the focus is on digital technology and the connections it facilitates, offline routines and established rhythms onboard ships, e.g. meal times, shift work, video games, gym activities and table tennis competitions, are all particular routine activities that help provide continuity and, hence, a sense of security for seafarers. Due to this, digitally facilitated, but fragmented connections might have both supportive and detrimental effects on security as it shapes and sometimes disrupts such routines. 

\subsection{Communicating at Sea}\label{sec:communicating-at-sea}
Existing work has highlighted the importance that seafarers attach to being able to communicate with home while they are at sea, separated from family and friends. For example, in a 2003 study based on qualitative interviews with seafarers' partners in China, India and the UK, Thomas et al.~\cite{MPM:ThoSamZha03} concluded that separation from family was the main cause of stress for seafarers. In a recent study, Oldenburg and Jensen~\cite{JOMEH:OldJen19} found that ``good'' communication with people at home was critical to mitigate the stresses and pressures felt by seafarers while away from their families. In a digital context, Tang~\cite{ESS:Tang09,SRO:Tang10} explored the role of online support groups for seafarer partners, highlighting the tension between the empowering and disempowering effects of such groups~\cite{SRO:Tang10} as well as how group members shared feelings and experiences of separation~\cite{ESS:Tang09}. Moreover, Papachristou et al.~\cite{WMU:PapStaThe15} emphasised the importance of communication with families in the retention of seafarers. These studies testify to the critical role that communication, both online and offline, plays in connecting seafarers and their home.

More broadly, existing research tying into aspects of digitally facilitated communications within maritime communities has predominantly focused on the effects on the mental wellbeing of seafarers and their families. For example, while focusing on transnationalism, Acejo~\cite{JIS:Acejo12} found how seafarers communicated, through mobile and satellite phones as well as emails, to maintain relations with family members. Swift and Jensen~\cite{SwiJen16} point to both positive and negative effects of digital communications onboard ships. How digital connectivity might limit the effects of social isolation at sea has also been highlighted by existing research~\cite{IMH:MelCar17,LTB:Trout80}. For example, Sampson~\cite{IMH:Sampson03} called for further research into the impact of social isolation on seafarers and accidents at sea, while Collins and Hogg~\cite{ColHog04} focused on the use of digitally enabled communication through, e.g. emails, to ensure seafarer welfare. From a security perspective, publications have been dominated by a focus on `maritime cyber security', with technological security being at the forefront of such research, see e.g.~\cite{DiGoRo15,TamJon18,WoGiHo17} and the references therein. This focus on technological security, while important, largely ignores the relational and social aspects that contribute to notions of security, as emphasised below. 

A number of industry reports and publications have also drawn attention to the need for (online) communication access for seafarers. While these are mainly based on survey data, they clearly exemplify the limited and fragmented nature of onboard communications. They show that although most seafarers have some Internet access, this is often restricted in terms of data allowance, accessible content and types of connectivity, e.g. no streaming, no video calls and no downloads~\cite{Nautilus17}. However, while still painting a fragmented picture of connectivity, research by the Seafarers' International Research Centre~\cite[p.39]{SIRC:SEATT18} noted a ``significant improvement in relation to Internet access for seafarers on board cargo ships in the period 2011-16,'' with 61 per cent reporting no Internet access on board in 2011 compared to 49 per cent in 2016. However, it also revealed that almost half (46 per cent) of the respondents were dissatisfied with the speed of their connection. While 93 per cent responded that they had access to text based chats, only 44 per cent and 61 per cent, respectively, could use video and audio chats. 

This survey data highlights that for seafarers being ``connected'' does not come without problems and challenges -- and it is not straightforward. For example, the Seafarers Happiness Index~\cite{MissiontoSeafarers19} shows that while seafarers ``crave'' connectivity, they also value time spent ``unconnected'', engaging with fellow crew members. And while 53 per cent of almost 6,000 seafarers responding to the 2018 Futurenautics survey~\cite{Futurenautics18} said they believed connectivity had reduced crew interaction, an increase in reported connectivity was matched by a three per cent fall in usage, suggesting that if there is a correlation between connectivity and seafarers' social isolation, it does not continue beyond a certain point. Moreover, 95 per cent of respondents considered connectivity to have improved safety onboard ships~\cite{Futurenautics18} -- which counters oft-heard concerns about connectivity compromising safety and security. 

While these reports testify to the rate of digital connectivity at sea, they do little to understand associated practices and notions of security that come with (fragmented) connectivity. In order to bridge this limitation of existing research, the subsequent sections bring them into conversation with situated HCI scholarship on the one hand and notions of security as lived experiences on the other. 
  
\subsection{Digital Exclusion}\label{sec:digital-exclusion}
Scholarship within HCI has discussed digital exclusion and disruption from multiple perspectives, often related to a lack of technological access, functionality and/or control, as highlight by e.g.~\cite{SDT:FrLiAh16,SDT:MouBra18}. Moreover, an emergent body of work within HCI sits at the intersections of digital exclusion and notions of security. In particular, research on marginalised groups has shown how an inability -- rooted in technological, social, economic or geographical constraints -- to digitally connect with essential services and social ties can have a direct impact on people's notions and feelings of security. This is particular evident in HCI work related to refugees and their security needs when interacting with digital systems~\cite{CHI:ColNew18,ColJen19,SP:SLIRK18}, but also geographically remote and isolated communities, e.g~\cite{pinto2018lora,CHI:WenJenCol20}. A growing body of work within HCI has also focused on how marginalised groups use digital technologies to break isolation, e.g.~\cite{davis2018including,johnson2011traffic,taylor2018remote}. Other situated HCI research explores how social relations are used to respond to technological security issues. Vines et al.~\cite{HFCS:VBDVTMO12} foreground the socio-materiality in relation to feelings of security in the context of financial security, Ogbonnaya-Ogburu et al.~\cite{CHI:OITD19} show how feelings of insecurity shape online practices, while Dourish and Anderson~\cite{HCI:DouAnd06} situate security questions within social and cultural contexts. 

Similarly, research on the importance of digital connections within marginalised groups has shown how the mobile phone is central in extending and securing access to social networks~\cite{CHI:ColNew18}, while others have pointed to a ``digital vacuum'' experienced by, for example, refugees in their digital interactions~\cite{CHI:JenColTal20}. Moreover, scholars have explored ways of designing for what Davis et al.~\cite{davis2017designing} call ``the digital fringe'' (see also~\cite{hespanhol2018digital}). The ``fringe'' in this context refers to communities with limited access to resources and digital technologies as well as people who are socially excluded or marginalised and, as a result, vulnerable to digital under-participation~\cite{davis2017designing}. This is especially relevant for the seafarer context, where limited and fragmented connections, mobile living and transitions between being connected and disconnected are inherent features of daily life and living. Hence, the present study should be understood within this growing body of work in situated HCI scholarship on digital exclusion and the implications for individual and collective security notions and practices.  

\subsection{Notions of Security}\label{sec:notions-security} 
Security when understood in relation to digital connections is often focused on the protection of the technology and/or of information, rather than on the protection or empowerment of people. However, understanding security as both enablement and protection, what Roe~\cite{roe2008value} calls positive and negative security, enables an interpretation of security that takes the routines and patterns of practice that make up people's daily lives as its point of departure and analytical lens. This understanding of security in the study of digital connectivity has been explored in recent work focusing on sociotechnical aspects~\cite{HAI:ColHan17} and, as mentioned above, in an emergent body of HCI and security research with marginalised groups such as refugees~\cite{CHI:ColNew18,ColJen19,CHI:JenColTal20,SP:SLIRK18}. In a seafaring context, established patterns of work and rest, including contact with family and friends as well as scheduled port visits and work routines, are part of seafarers' daily working and living. These are repetitive actions that form a daily routine for many and, as a result, provide a sense of normalcy that support feelings of security. 

This understanding of security is tied to the basic notion of ontological security~\cite{CUP:McsSec99}, where security comes from a sense of each being secure in the other. In other words, here, the notion of security is established through trust bonds within relations and interactions~\cite{giddens1991modernity} and is often supported and strengthened through established routines and regular patterns of practice. From this perspectives, routines are important for maintaining a sense of security~\cite{UCP:GidCon84,roe2008value} because they create feelings of continuity that connect individuals both to the environment in which they exist but also to the relationships that they have within and beyond that environment. This includes relational ties with and within broader social networks, cultivated and maintained through both online and offline interactions, as well as established patterns of practice. From this perspective, for the seafarers in this study, onboard routines relating to meal times, work shifts and recreational activities all create continuity and patterns of work and rest that help establish security, while these are often disrupted by the fragmented patterns of digitally facilitated communication. This study brings this notion of security into conversation with the seafaring context, through an ethnographic research approach, as outlined below. 

\section{Research Methods and Analysis}
Ethnographic in nature, the study was designed to engage with seafarers where they live and work; onboard ships. This speaks to previous ethnographic work with seafaring communities, e.g. Baum-Talor~\cite{baum2014sea} provided an ethnographic account of life onboard merchant ships, Sampson~\cite{sampson2013international} explored transnationalism on ships through onboard fieldwork, while Galam~\cite{galam2015gender} reflected on gender and positionality in working with seafaring families. The present study engaged 43 seafarers onboard two large container ships, during two two-week voyages in European waters in February and April 2018 (see Table 1); one with onboard WiFi facilities and one without (see Table 2). All participants were experienced seafarers, who had spent most of their lives at sea -- except for the two deck cadets. The ships were modern and had spacious living accommodation, recreational and socialising spaces and entertainment facilities, including pre-recorded local news media, television series and movies. 

\begin{table}[h]
	\centering
	\begin{tabular}{lll}
		\hline
		&\textbf{Ship 1}&\textbf{Ship 2}\\
		\hline
		\textbf{Location}&European waters & European waters\\
		\textbf{Timings}& 14-30 Mar 2018& 2-16 May 2018\\
		\textbf{Participants}&22  & 21 \\
		& (8 officers, 14 crew) & (8 officers, 13 crew)\\
		\textbf{Gender}&Male& Male\\
		\textbf{Age}&23-57&21-59\\
		\textbf{Nationality}&Filipino, Ethiopian, & Filipino, Chinese\\
		&Indian, Sri Lankan, & Indian, Sri Lankan,\\
		&Ukrainian & Ukrainian\\
		\textbf{Working language}&English & English\\
		\hline
	\end{tabular}
	\caption{Research design and structure for the seafaring case study.}
	\label{TAB:ResDes-sefaring}
\end{table}

\noindent Access to ships was secured with the assistance of a large shipping company. This enabled one female researcher to conduct empirically grounded research during two voyages in European waters. The researcher embarked in London and Rotterdam, respectively, and disembarked in Piraeus. Both voyages included several port stays in Hamburg, Antwerp and Le Havre, which each lasted between 12 and 36 hours. The study adopted an unobtrusive methodology designed to work around routine tasks and schedules. Group discussions and conversations were arranged in non-formal setups (e.g. during meal hours, socialising and between work shifts) and the researcher was able to observe all aspects of work tasks.

\begin{table}[h]
	\centering
	\begin{tabular}{lll}
		\hline
		& Ship 1 & Ship 2 \\
		\hline
		\textbf{WiFi}& Yes & No\\
		\textbf{Data }& 50MB pp per week (free) & N/A\\
		\textbf{Email}& Individual emails & Individual emails \\
		& accessed through  & accessed through  \\
		& shared computers  & shared computers \\
		& or mobile phone app & or mobile phone app \\
		\textbf{Phone}&Calling cards (paid for)& Calling cards (paid for)\\
		\textbf{Media}&Shared TV and computer & Shared TV and computer\\
		& in common rooms,  & in common rooms,  \\
		&  access to local news, &  access to local news,\\
		&  movies and TV series &  movies and TV series  \\
		\hline
	\end{tabular}
	\caption{Onboard access to media and communications technology.}
	\label{TAB:connectivity-sefaring}
\end{table}

\paragraph{Ethics.} The study was approved by the research ethics committee at the researcher's institution. Participation was voluntary and contributions were anonymised through the analytical process. Participants received individual information sheets, which explained the purpose of the study and the researcher's presence on board the ships. This happened on the first day of each voyage, but both ships had also received this information before the researcher embarked. Participants were also given an opportunity to ask questions before signing individual consent forms. It was made clear that their decision to contribute or not would be kept confidential, as would their contributions. All 43 seafarers (22 on Ship 1 and 21 on Ship 2) were keen to contribute and signed the individual consent forms. As one crew member on Ship 1 notes: \emph{``We've had a survey, and no-one will say `no' to more Internet [\dots] but it's not that simple. The people creating the survey don't understand what it means not to have good Internet. Because you're here, you will experience it.''}

\paragraph{Data capture and analysis.} Three forms of data were captured during the study: (1) written notes from group discussions, which were captured verbatim where possible; (2) researcher observations captured in note form (field notes); and (3) images of work and life onboard both ships, captured by the researcher. The data analysis was inspired by Gillian Rose's analytical approach~\cite{rose2016visual}, which enabled qualitative interpretations of both visual and textual data. Systematically coding and interpreting these data in relation to the seafarer context revealed a series of categories, including mobility, health and safety, stress and fatigue, work pressures, family pressures, rhythms and routines, on-board camaraderie, loneliness and homesickness. The categories presented here focus on the routine everyday tasks and pressures related to social relations, and their intersection with digital technology and notions of security. 

\paragraph{Limitations.} The study was based on data collected from fieldwork on two container ships owned by the same company. Findings might have provided a more nuanced picture if fieldwork had been conducted with crew onboard ships owned by companies that allow more extensive WiFi access on their ships. However, none of the participants in this study had ever experienced unlimited or reliable online access, although they had worked for different companies and charterers. It is also possible that the study was shaped by the European setting within which it was conducted, while most participants were from countries in Asia. Their experience of being out-of-sync with their social networks, e.g. because of time differences, might therefore also have been felt to a greater extent. Nevertheless, the data shows a clear set of categories as well as many and varied security experiences articulated by participants. Future research should aim to expand the contextual settings and engage with seafaring communities more broadly; this includes doing fieldwork in ports, especially in Seamen's Clubs where seafarers often connect with home, and with seafarer families. 

\section{Research Findings}\label{sec:research-findings}
This section outlines the research findings that emerged through the analysis of the ethnographic field data, focusing in particular on two overarching categories: security through routines in Section~\ref{sec:security-routines} and disrupted connections in Section~\ref{sec:disrupted-connections}. In presenting the findings, this section employs the `ethnographic present' to convey the lived experiences of the seafarers who took part in this study.\footnote{The `ethnographic present' is discussed in several works in anthropology in particular, see e.g.~\cite{SA:DePina-Cabral00,CA:Hastrup90}.}  

\subsection{Security through Routines}\label{sec:security-routines}
The study reveals how notions of security take many forms onboard ships and how daily, habitual routines are often structured around negotiations of social relations related to fragmented connectivity. This fragmentation is evident in all of the data. The data also brings to the fore the extent to which recent changes to seafaring has further limited the ability for seafarers to stay connected with their social networks. For example, less time in ports and limited shore leave have reduced access to WiFi facilities in ports, while reduced ship speed has meant that seafarers spend longer time at sea often without satisfactory online access. Combined with stricter socialising and alcohol consumption policies, and larger ships with smaller crews, a limited ability to stay connected with home has led to feelings of isolation and exclusion. This is a recurring observation in the data. This also shows how the disruptions created by fragmented digital connections are not, in most cases, mitigated by onboard socialisation. Moreover, seafarers highlight how the ``digital divide'' between the ship and their home makes the lack of reliable digital connectivity onboard ships particularly unsettling. While previously, they have been used to being disconnected from their families for extended periods of time, now: (1) they know that Internet access is a possibility and therefore not having it becomes a disrupting factor in its own right, (2) their families are digitally connected and often expect to be able to communicate regularly and (3) the transition between being at home and being at sea is seen as more difficult due to this digital divide between the two contexts. 

For seafarers, such changes to seafaring, as well as to their external environment, has a significant impact on their daily routines and ability to feel in control of such routines, which also affects their notion of security. This is despite the fact that all participants, except for two deck cadets, were experienced seafarers and had spent most of their lives at sea. They highlight how feelings of being isolated, excluded and disconnected from \emph{``the rest of the world''}, partly due to such changes to their work environment, make them \emph{``crave''} other forms of connectivity. Moreover, the fact that \emph{``the rest of the world''} is now able to connect through digital technology, makes many of the seafarers in this study feel even more excluded from familial and social contexts. 

The researcher embarked when each ship arrived at its first European port. Prior to this, both ships had been to several ports in Asia. One crew member on Ship 2, which had no on-board Internet, notes: \emph{``The voyage from Singapore to Rotterdam is the worst [\dots] we don't go into port and are completely isolated from the world for three weeks.''} The crew on Ship 1 also stress how their 50MB weekly data allowance means that they have almost no contact with people at home during long sea passages. The priority for everyone on both ships is to buy a European SIM card when they reach Europe: \emph{``When we arrive in Rotterdam, we're desperate. And if the SIM card seller doesn't show up in port immediately, everyone's sad''} (Ship 2).

In these moments, when connectivity is restored, the seafarers speak of the relief, pleasure and happiness they feel at being in contact. The movement of the ship, in and out of port at specific times, means that such moments become routine practices, which enable them to manage and negotiate their fragmented existence between being connected, not connected or in between. One of the younger crew members notes how being able to see his newborn baby, who he has not yet physically met, through FaceTime makes him feel \emph{``like I can make it through another week''.} This is seen as critical for the seafarers as they experience closeness and distance in differentiated ways, shaped by their everyday lives onboard ships, whilst intimately connected with life at home. \emph{``We might be physically distant but we're not distant emotionally,''} as one crew member on Ship 1 notes. 

This kind of co-presence is further distorted and sometimes disrupted by uneven and unreliable digital connections. On Ship 1, all crew members explain how they \emph{``budget''} their 50MB weekly data allowance. While some make their data last by not exceeding 7MB per day, others save their data to use on one particular day. The primary purpose for connecting is to communicate with family, mainly through WhatsApp and Imo, as other services, such as Facebook and Instagram, are seen to \emph{``consume too much data''}. Group discussions as part of the study reveal how seafarers ration their Internet usage by using low data consumption applications or by structuring their work and rest routines to connect when the ship is in phone signal range:  \emph{``Sometimes you don't want to go to sleep because there is a chance that you will have a signal''} (Ship 1). This needing to connect every time the opportunity arises often perturbs the routinisation required to maintain a sense of security, established through continuity of daily, established tasks and movements. Several examples emerge from the ethnographic data of how this leads to limited rest between work shifts. Many of the seafarers comment that they value having regular contact with home more than sleeping and will set an alarm if they think that there is an opportunity to connect.

Onboard observations also reveal that security gained from the routine of managing social relations by rationing data usage or by trying to predict when the ship is within phone signal range is often disrupted by changing sailing schedules: \emph{``You may have planned to message someone or speak to your family when you're in a certain port on a certain date, but when the schedule then changes, these plans are disrupted and you feel alone''} (Ship 2). Several examples are given where crew members have missed a child's birthday or a friend's anniversary or wedding due to unforeseen delays or changes. Being physically separate, while having to negotiate uneven and fragmented digital connections, intimately shapes feelings of insecurity among the seafarers in this study. The routines of communicating with home are therefore often broken or never fully established. Because of this, while seafarers note how the ability to be able to regularly connect with family, as a way of establishing routines in this regard, help them feel more in control, the fragmented and unreliable nature of their digital connections makes this impossible. 

\subsection{Disrupted Connections, Pressures and Uncertainty}\label{sec:disrupted-connections}
The study highlights how the reworking of seafarer lives, increasingly interwoven with fragmented digital connections, creates a series of pressures. This is also evident through the multiple restrictions, limitations and costs which all contribute to the complex picture of multi-layered connections running through, within and beyond the boundaries of the ship. Although participants onboard Ship 1 note that limited connectivity is better than no connectivity, there is general consensus that this connectivity is not enough to maintain satisfactory and intimate relations with wider social relations and networks, especially during long periods at sea. In particular, these limitations mean that it is not possible to use Skype or FaceTime, download videos or music, or communicate beyond sending text messages. Data allowance is therefore only seen to be useful for sending brief messages through applications, such as WhatsApp or Imo, which, however, will need to be limited and rationed. This is seen to generate a lot of uncertainty and frustration amongst seafarers. For them, \emph{online} does not simply mean \emph{online} and \emph{connected} does not simply mean \emph{connected}, but it comes with a number of pressures and stresses, and it refers to a myriad of networks, connections and relations that are all interwoven in the seafaring context.

Financial pressure in particular dominates the data in relation to digital connectivity. Participants speak of differentiated salaries between crew members depending on rank and nationality and how this impacts on their ability to stay connected. For most of the crew, it is not an option to buy SIM cards in every port: \emph{``It's fine for officers to spend a lot of money on phone cards, but it's different for ratings.''} Instead, they weigh up the value of spending money on a SIM card in one port over the other. Crew members on both ships also note that they are prime targets for scammers who are selling SIM cards that do not contain either the minutes, speed or data that they are promised. These SIM card sellers are described as \emph{``the Mafia''}: \emph{``in most ports, SIM card sellers come onto the ship to sell their stuff. We call them 'the Mafia' because they cannot be trusted but we're reliant on them for connectivity.''}\footnote{The study also suggests that these types of scams usually happen in some ports in Africa, South America and Asia, but very rarely in Europe, America, Canada, Japan, Singapore, Australia and New Zealand. This is due to strict port regulations.}

Another pressure in relation to digital connectivity expressed by many seafarers in this study is the pressure coming from family members, who often find it difficult not being able to stay digitally connected during separation. It is unclear from the study how much the seafarers themselves know about the connectivity available on a given ship before embarking. Many seafarers note that they have to manage family members' expectations about access to digitally facilitated connections when they are away, however, that this is difficult since they do not know themselves. This was previously not an issue since digital technology was not as prevalent in societies more broadly. 

Furthermore, seafarers' ability to connect with people at home on a regular basis is understood to ease transition into home life when returning from up to nine months at sea. In this phase of transition, being in frequent contact allows them to keep up to date with events and activities at home. This sense of being \emph{``in control''} of their lives and not \emph{``excluded''} helps them maintain a sense of routine, which also impacts their security. The data also reveals that being in control contributes to the maintenance of resilience during long periods at sea. Yet given the fragmented and disconnected digital connections onboard both ships, seafarers spoke of \emph{``being out of sync with home life''}, which, in contrast, contributes to individual feelings of insecurity. 

The study demonstrates how fragmented connections, which surface when the ship moves in and out of connectivity or when onboard data allowances run out, impact on individual and collective feelings of security. These mundane experiences of how, on the one hand, routines are disrupted by fragmented digital connectivity and, on the other, the inability to establish reliable routines through digital connectivity, e.g. through regular contact with home, perturb feelings of security among seafarers. 

\section{Discussion}
The study highlights how notions of security and the seafarer context are interwoven and shape digital interactions -- and vice versa. In particular, the fact that seafarers often have to rely on uneven and unreliable digital connections to stay in touch with their wider social relations, during long periods at sea, has been shown to disrupt daily routines and security. This study has exemplified the differing nature of connectivity onboard ships and how fragmented spaces, between connected and disconnected, perturb daily routines needed for seafarers to establish a sense of security. In this section, suggestions as to how the maritime community might consider future approaches to technology design and practice are discussed. 

Evident from Section~\ref{sec:research-findings}, for the seafarers in this study, notions of insecurity are often connected to fragmented onboard digital connectivity. This is largely due to the fact that it underlines their isolated existence, often further emphasised by the digital divide between the ship and their home -- as a result, heightening their sense of being digitally excluded. This divide limits their ability to routinely connect with their family and friends. Importantly, however, a number of other routine tasks and daily occurrences on board the ships also have a direct impact on security. As noted above, established routines related to meal times, work shifts and recreational activities all create continuity and rhythm that help build emotional resilience and a sense of security. However, given the importance that seafarers attach to being able to digitally connect, the fragmentation of these connections are seen to disrupt these other routines. 

The particularities of the seafaring context exemplify how increased, yet uneven and unreliable digitalisation can disrupt established routines. On the other hand, the study also shows how the efforts undertaken by seafarers to re-establish routines are both extensive and varied and are focused on negotiating the social exclusion that emerges in the intersection between connectedness and separation. This presents an uncertainty that some seafarers in this study try to overcome by staying digitally connected with their wider social networks and relations; however limited such connections are. On the other hand, some participants have tried to ``switch off'' to avoid what they see as the negative impact of uneven connectivity. This is often a practice highlighted by more senior seafarers, who have had extensive experience of being at sea before digital technology and Internet access had emerged as key expectations. For younger seafarers in particular, being digitally, often unevenly, connected with home, whilst physically disconnected (being physically absent whilst emotionally present), is generally seen as more beneficial than not being connected at all. 

The study contributes nuanced understandings of fragmented connections. It shows, albeit differently, how attempts to (re)establish routines within the seafarer context, are, in some ways, responses to notions of (in)security. And where routines are either broken or fragmented, they can have a detrimental affect on an individual's security. In this work, I argue that this needs to be considered by the maritime community when designing policies and practical interventions that aim to minimise the potentially negative effects of living and working at sea.

\subsection{Practical Implications}\label{sec:practical-implications}
One of the central arguments within the maritime industry for not providing unlimited onboard Internet access has been that it would disrupt work and rest patterns onboard ships, which could ultimately compromise safe and efficient ship operations~\cite{Nautilus17}. However, this study shows that, in fact, not having reliable Internet access onboard ships significantly disrupts such patterns. The study exemplifies that if the only method of connecting with broader social networks and with home is through personal mobile phones, seafarers will do so when the ship is within mobile phone signal range, regardless of the time of day, external factors, work or rest hours. This can ultimately have a direct impact on general welfare and wellbeing of seafarers, who negotiate and manage several digital connections, at any one time.

It is critical to recognise that digitally facilitated connections have become essential to the lives of seafaring communities and, therefore, should be an integral part of modern seafaring. Regardless of whether shipping companies provide onboard Internet access, seafarers navigate and negotiate several interwoven digital connections every single day. They do so largely to minimise emotional stresses of being separated from their families, to establish a sense of security and to build emotional resilience. This might be done by circumventing access restrictions, sharing account passwords, buying mobile phone SIM cards in ports, rationing data use, and monitoring ship positions to predict when a mobile phone signal will be available. Digital connectivity at sea therefore needs to be at the forefront of current thinking, future policy discussions, ship and technology designs.

The study shows that limited onboard connectivity and the pressures linked to finding ways of connecting amplify other pressures related to work, isolation, separation, family and finances. This is exemplified by the buying and selling of mobile phone SIM cards in ports. Seafarers feel that they have \emph{``no other choice''} than to spend money on these SIM cards, even in places where they know that they are prime targets for potential scammers. Furthermore, access to onboard connectivity is increasingly becoming a deciding factor whether `young and talented individuals' want to spend their lives at sea, and whether shipping companies are able to retain experienced and highly qualified seafarers, see e.g.~\cite{MartimeLogistics20}.

Approaches focusing on technological innovation alone, however, are unlikely to be successful within seafaring communities. It is therefore necessary for the maritime sector to consider the social, cultural and economic aspects within which digital technology is situated and which have been shown to have a direct and indirect impact on seafarers' notions of security. 

In line with this, the study suggests an approach driven by a number of principles aimed at ensuring that technologies and practices related to digital connectivity counter the uncertainties experienced by seafarers. First, providing free and reliable onboard Internet access should be central to any policy discussion. While this does not solve all challenges facing the seafaring community, instant, regular and stable digitally facilitated contact with home helps alleviate pressures, emotional stresses and build emotional resilience amongst seafarers. These are critical for seafarers to build and maintain security. Second, constantly moving in and out of digital connectivity disrupts sleep and work routines on the one hand and creates fragmented connectivity on the other. This study has indicated that this can have a negative impact on feelings of security and seafarer resilience. More should be done to even out the unevenness of digital connectivity onboard different ships and across shipping companies. Third, active collaboration between seafarers, shipping companies, trade unions and Internet service providers is needed to share ideas and co-design potential solutions with an aim of creating interventions and/or technologies that reflect the security needs of seafarers. Finally, more research should be done in wider maritime settings to provide nuanced understandings of the impact of digital connectivity onboard ships on seafarer's lived security experience; to facilitate security discussions that do not take technological security as their point of departure.

This study should be seen as a starting point for a wider programme of engagements with seafaring communities, focusing on their sense of collective and individual security in relation to digital connectivity, in order to further demonstrate the critical importance of reliable Internet access for seafarers.

\section{Conclusion} 
With these practical implications in mind, the study concludes that in order to support the seafarers and to enable them to build individual and collective resilience, access to free and reliable digital connections and networks is critical. This not only helps minimise disruptions to daily routines and patterns of practice, but it also counters the unevenness of digital connectivity onboard ships and across shipping companies. While this does not solve all challenges facing the maritime industry -- nor does it ensure seafarer wellbeing and onboard cohesion -- instant, free and stable digitally facilitated interactions are critical to counter the uncertainties emerging from disrupted and unreliable connections. Interpreting the research findings through this particular security lens highlights the significance of digital connectivity in seafarers' notions of security. This enables technological initiatives to be brought into conversation with individual feelings of security. Future practical and policy interventions need to consider the relationship between the two in order to reduce feelings of isolation and exclusion. Finally, digital connectivity cannot and should not be understood as separate from the social environment within which it exists. For seafarers, this environment is fragmented.

\section*{Acknowledgements}
I am particularly grateful to the research participants for taking part in this study. Without their efforts, enthusiasm and energy this work would not have been possible. For comments and suggestions, I would like to thank the anonymous reviewers and the editors of this special issue. The research was supported by the Sailors' Society.


\bibliography{refs}


\end{document}